# PROMPTING WHISPER FOR JOINT SPEECH TRANSCRIPTION AND DIARIZATION

*Mariia Zamyrova, Henk van den Heuvel*

Radboud University Nijmegen

**ABSTRACT**

As part of the MediSpeech project, we aim to develop a system that transcribes and diarizes Dutch conversations between doctors and patients in real-time. In this research (in-progress) we explore ways of efficiently combining Whisper with speaker diarization (SD). After trying to prompt Whisper with text that contains speaker labels, we observed that it is able to insert labels into the transcription with promising accuracy. We continued this line of research by fine-tuning Whisper with speaker-labelled prompts to generate transcriptions in a format similar to that of Serialized Output Training (SOT). Fine-tuning Whisper yielded more consistent speaker IDs across the chunks of long-form audio and improved verbatim transcription. The study uncovered new challenges as Whisper's SD performance suffers because of mistakes that get propagated through prompts and inaccurate timestamps assigned to overlapping speech.

## 1. INTRODUCTION

Whisper [1] has established itself as a front-runner when it comes to automatic speech recognition (ASR). A feature that particularly makes Whisper stand out is its prompting functionality. By using text prompts, we can condition Whisper's transcriptions, which can be helpful for transcribing chunked long-form audio or correcting domain-specific terminology spelling. Prompts have also shown to be useful for adapting Whisper to new tasks. For example, previous research used ground truth text as prompts for Whisper to train it to detect reading miscues [2].

Inspired by the versatility of prompting, we wanted to use it to adapt Whisper to our use-case: real-time joint speech recognition and diarization for Dutch. We observed that prompting Whisper with text that contains speaker labels, e.g. "[Spreker 1] Hallo! [Spreker 2] Hallo! Hoe gaat het met jou?" (Eng.: "[Speaker 1] Hello! [Speaker 2] Hello! How are you?"), enables the model to insert speaker labels into the transcription with promising accuracy, albeit with varying generalizability and speaker label hallucinations. In the last couple of years there has been significant progress in embedding diarization into the Whisper ASR pipeline [3-6]. However, challenges remain for use in real-time scenarios. Some approaches require a speaker-specific diarization mask or a target-speaker embedding and thus process each speaker individually which can increase computational complexity for multi-speaker audio [3-4]. Additionally, Serialized Output Training (SOT) [7] approaches do not consider long-form inference and do not preserve the same speaker IDs across chunks of audio [6]. The possibility of injecting each speaker's labels into Whisper output by only using prompting is exciting because of the lowered computational costs. Conditioning long-form audio chunks on speaker-labelled prompts could also help preserve the speaker IDs throughout the entire audio file. Thus, prompting Whisper could offer a solution to both of the above challenges.

In this research we fine-tune Whisper using a speaker-labelled prompting scheme. Through adapting the model to a single prompting scheme we strive to remedy the previously mentioned generalizability and hallucination issues. To ensure that the model reuses the same speaker labels throughout chunks of a long-form audio we provide it with both a "task prompt" that contains speaker labels that we want the model to choose from and a "context prompt" – a transcription of the previous audio chunks from which the model should infer if there was a speaker change and what speaker label follows next. With this we aim to create a model that simultaneously performs transcription and diarization (in real-time) and is able to combine textual context and audio information to generate robust speaker labels. Our results showed that while the diarization performance is varied (especially for 3+ speakers), prompt-tuned Whisper is indeed capable of generating uniform and continuous speaker labels for long-form audio, and, in addition, is better at verbatim speech recognition.

## 2. DATA

We make use of the Corpus Gesproken Nederlands (CGN)

| # speakers | Train | Valid | Test | Test subset |
|---|---|---|---|---|
| 2 | 503 | 53 | 121 | 14 |
| 3 | 106 | 29 | 65 | 4 |
| 4 | 35 | 10 | - | - |
| 5 | 3 | - | - | - |
| Total duration | 99.5 | 12.5 | 27.0 | 2.8 |
| Avg duration | 6.9 | 6.6 | 7.3 | 8.5 |

Table 1: For each partition of the CGN comp-A dataset: the number of audio files with N-speakers, the total amount of hours of speech, and the average duration of speech in a file in minutes.

dataset [9-10], specifically the "comp-A" part of the dataset that contains 925 long-form recordings of spontaneous face-to-face conversations (Table 1). The conversations mostly center around general topics, such as discussing movies. The component contains recordings from 231 participants, all native "Standard Dutch" speakers. All of the speakers are older than 18 with 56.7% being 18-34 years old. The dataset includes speakers of both sexes with 55% female and 45% male speakers; 59.3% of audio files contain both male and female speakers. The amount of overlapping speech in each audio varies, largely making 5-10% of the whole duration of speech. The majority of overlaps span under 1 second, however there is a small portion of over 3 second overlaps. The audio files are broken into time-aligned utterances, each containing a single speaker. Each file is paired with an orthographic transcription, with speech artifacts like interjections, self-corrections and broken-off words preserved in the text.

As there is no official partition of the dataset we split it in a 7:1:2 train, validation and test ratio, respectively. We also use a ~3 hour subset of the test set for prompt engineering and more detailed evaluation of the results. We note that the speakers in the test set are not present in the train or validation sets.

## 3. METHODS

### 3.1. Prompting

After more experimenting we found that prompting the model with just the speaker labels as hotwords, e.g., "[Speaker 1] [Speaker 2]", yielded a labelled transcript similar to the one we got with a full sentence prompt while taking up fewer tokens. During training, to further minimize the size of the prompt, we replace the labels with "[Sn]", where n $\in$ [1,5] as we saw that there are at most 5 speakers per audio (Table 1). We feed the model all of the speaker labels at once ("[S1] [S2] [S3] [S4] [S5]") rather than adjusting them to the (known) number of speakers in each sample. This way we aim to create a more flexible model that infers which speaker labels it needs to use and does not rely on the number of speakers being known in advance.

To simulate long-form inference conditions during training and enforce consistent speaker IDs throughout chunks of the same audio we generate the prompt for each training and validation sample in the following way:

- We group the utterances belonging to the same audio file and sort them chronologically
- The first utterance receives only the "task prompt", i.e. speaker label hotwords, as there is yet no context
- Every consecutive utterance receives both the "task prompt" and the accumulated speaker-labelled reference text of the previous utterances as the "context prompt". For example, "[S1] [S2] [S3] [S4] [S5] [S1] Uhm moeten langs uhm de Gamma gaan denk ik voor uh uh om die…" (Eng.: "[S1] Uhm must stop by uhm the Gamma I think for uh uh so that…")
- The "context prompt" is left-side truncated to fit within the length limit (224 tokens for the prompt and 224 for the training label)

Another behavior we want to enforce is the first chunk of an audio always starting with "[S1]". We randomly select 20% of the utterances and provide them with only the "task prompt" irrespective of their position in the audio file. If the utterance's reference transcript did not start with "[S1]" we relabelled the speakers within the text accordingly.

### 3.2. Fine-tuning

We expected the decoder's cross-attention module to have the most influence on the prediction of the speaker labels, being the main point of interaction between the text and the encoded audio features. Thus, we try tuning only the cross-attention and the two linear modules that follow it for every decoder layer. We fine-tune with Low-Rank Adaptation (LoRA) [8] and use the standard cross-entropy training loss without masking the prompt tokens. We hypothesize that by not masking the prompt Whisper will learn a stronger semantic correlation between prompted context and output.

## 4. EXPERIMENTAL SETUP

### 4.1. Model

We use Whisper *large-v2* for fine-tuning and as our baseline. During tuning we use the *Huggingface* implementation and modify the model's 'config.json' file by removing all non-special tokens from the 'suppress_tokens' list. For long-form evaluation on the test set we use the quantized *Faster Whisper* implementations of the base large-v2 and our tuned model.

### 4.2. Training configuration

We wrap the Whisper model with the *PEFT* library LoRA (rank=16, alpha=16) wrapper. We use the *Huggingface Trainer* with the default AdamW optimizer (learning rate of 1e-5) as our core training pipeline. We train the model on a single NVIDIA RTX A4000 for ~3 epochs (2430 steps) with a batch size of 4 and gradient accumulation factor of 4. The training and validation sets are shuffled prior to training with a random seed of 42.

### 4.3. Data pre-processing

For tuning Whisper we merge the CGN training and validation utterances into chunks of at most 30 seconds. We filter out utterances that contain unintelligible speech (“xxx”) or laughter (“ggg”) annotations in their transcriptions. We did not merge utterances if their accumulated text exceeded the 224 token length limit. To inject speaker labels into reference transcriptions we map the speaker IDs used by CGN to our “[Sn]” labels based on the order in which the speakers appeared in the audio file. We insert a label at the beginning of every utterance and every time there was a speaker change within the utterance.

CGN utterances correspond to either complete sentences or are part of a sentence that spans multiple utterances. If an utterance does not start with a new sentence or does not end with a punctuation mark, we pad its text with “…” to imitate Whisper’s annotation style in such scenarios and avoid deteriorating the model’s sentence casing. Because the ground truth sentences are not capitalized we also apply regular expression-based capitalization. Lastly, we remove any meta-data annotations from the text, such as “*a” markers signifying broken-off words.

### 4.4. Evaluation

As our target domain is long-form real-time audio inference we test on full audios from CGN test set rather than individual 30s chunks. At inference, the tuned model uses the same prompting scheme as during training, with the exception that the “context prompt” is now the predicted text from the previous chunks, rather than the reference text. We use the *MeetEval* Python library that includes the standard word error rate (WER) as well as constrained minimum-permutation WER (cpWER) that finds the optimal alignment between the text of reference and predicted speakers [11]. To evaluate per-word errors, we use the aligner from the *kaldialign* library to find pairs between reference words and their deletion, hit or substitution in the prediction. Prior to computing the word error metrics we split the text by the speaker labels using regular expressions and normalize the new segments. The normalization procedure for both reference and predicted texts includes lowercasing, removing punctuation and converting digits to their written form. To account for hallucinations, we remove any annotations between “[]”, “()” or “*” signs that do not get picked up by our speaker label regular expression.

To measure diarization accuracy we used *pyannote-metrics* diarization error rate (DER) with a collar of 250ms to account for deviations in Whisper timestamps and possible deviations in the automatic reference timestamps.

| | **WER (%) ↓** | | |
|---|---|---|---|
| | **min** | **2nd quartile** | **3rd quartile** |
| **a)** | 21.2 | 36.7 | 40.7 |
| **b)** | 21.2 | 37.2 | 41.5 |
| **c)** | 20.2 | 85.6 | 94.5 |
| **d)** | **6.9** | **31.7** | **37.7** |

Table 2: WER for the model configurations described in Section 4.5. To characterize the variation in the models’ performance we report the minimum observed WER, the median (2nd quartile) and the 75th percentile (3rd quartile). We highlight the lowest values in bold.

### 4.5. Model configurations

To evaluate the effect of prompting on the overall accuracy of the transcriptions (their WER), we compare the following models:

a) the baseline untuned model;
b) untuned model with a prompt engineered on the test subset: “[Spreker 1] Hallo! [Spreker 2] Hallo! Hoe gaat het met jou? [Spreker 1] Goed, en met jou? [Spreker 2] Goed. [Spreker 1] Mooi weer, he? [Spreker 2] Ja.” (Eng.: “[Speaker 1] Hello! [Speaker 2] Hello! How are you? [Speaker 1] Good, and you? [Speaker 2] Good. [Speaker 1] Lovely weather, huh? [Speaker 2] Yeah.”);
c) untuned model prompted using the same approach as the one we use during fine-tuning;
d) the proposed prompt-tuned model.

We also compare the performance of the tuned model with and without voice activity detection (VAD) to investigate the effect of speech pauses on speaker change detection. Most reported results were computed with VAD unless specified otherwise.

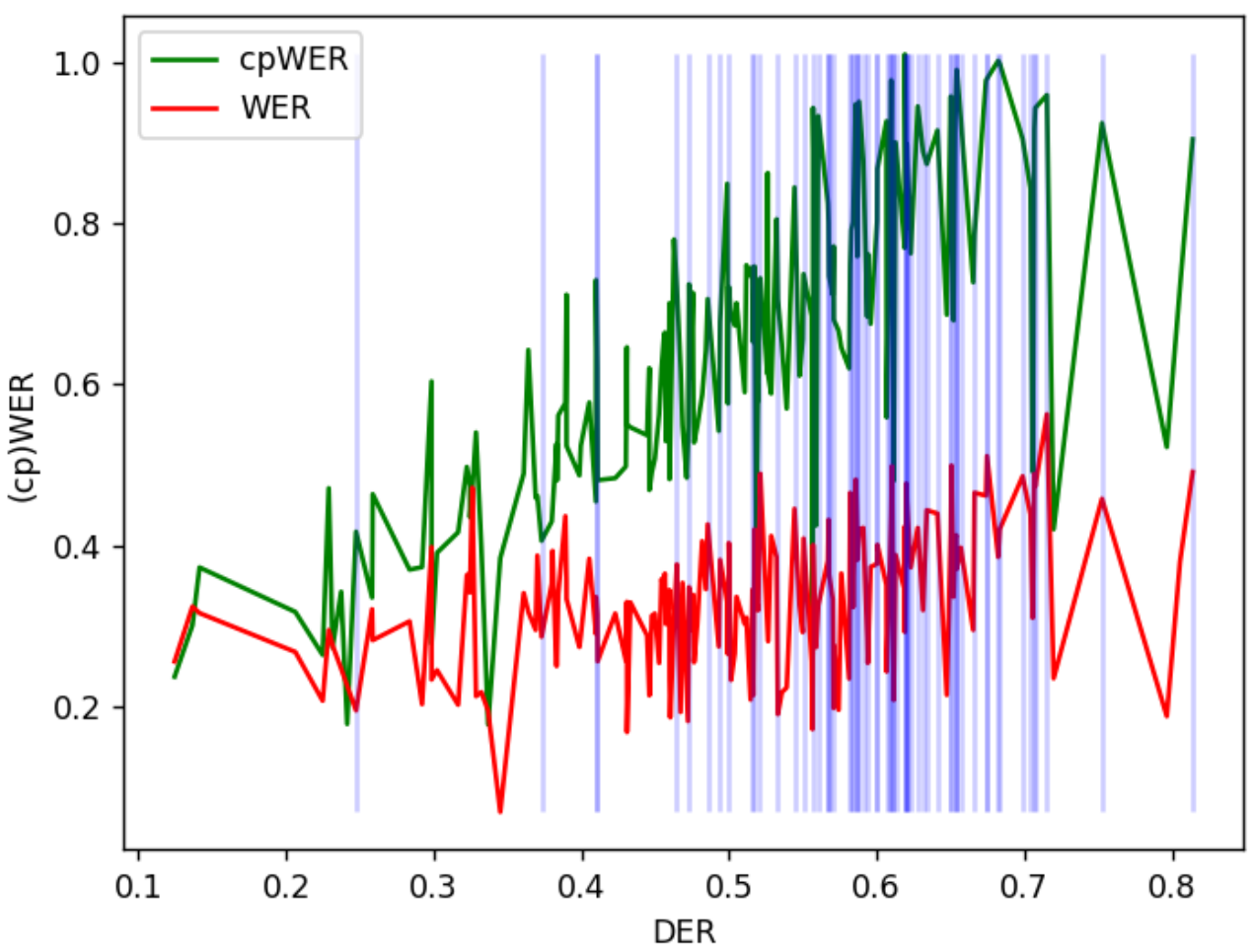

Figure 1: WER and cpWER plotted against DER. We shade the points that correspond to 3-speaker audios (in blue).

## 5. RESULTS AND DISCUSSION

One of our first observations after fine-tuning was the improvement in overall (verbatim) WER (Table 2). Especially, we found that the rate of filler word recognition increased from 7% hits to 63% hits for top 7 filler words with frequency above 500 over the whole test set (e.g., "uh", "oh" and "m"). This aligns with the results of [2] where the authors also observed improvements in verbatim ASR after prompted tuning. Bettering verbatim speech recognition offers many advantages particularly for our medical use case. For example, more robust performance on unseen medical terminology, especially given the challenges of accessing and collecting medical speech datasets.

We also saw that Whisper was able to adapt to our prompting strategy (Table 2, c vs d) that was not directly engineered on the test subset (unlike b). This is a positive sign for our future work that we can adapt Whisper to an arbitrary prompt without detriment to WER or large tuning expenses. Lastly, we can see that the WER is generally quite high for all model configurations. This is in part due to CGN only providing orthographic transcriptions that deviate from standard word spellings to which Whisper gravitates. Moving forward, we aim to standardize the text to reduce the variation in some of the spellings before fine-tuning and evaluation.

Regarding the model's speaker labelling, we found the performance to be mixed. Compared to the performance on the engineered prompt, the tuned model generates significantly fewer out of vocabulary (OOV) speaker labels. The untuned model created audio event labels such as "[Speaker 2 laughs]" or used speaker names like "[Judith]", with over 10 OOV labels (2% of all assigned labels). The tuned model had 2 OOV labels, "[Rudy]" and "[J]", which made up 0.02% of all labels.

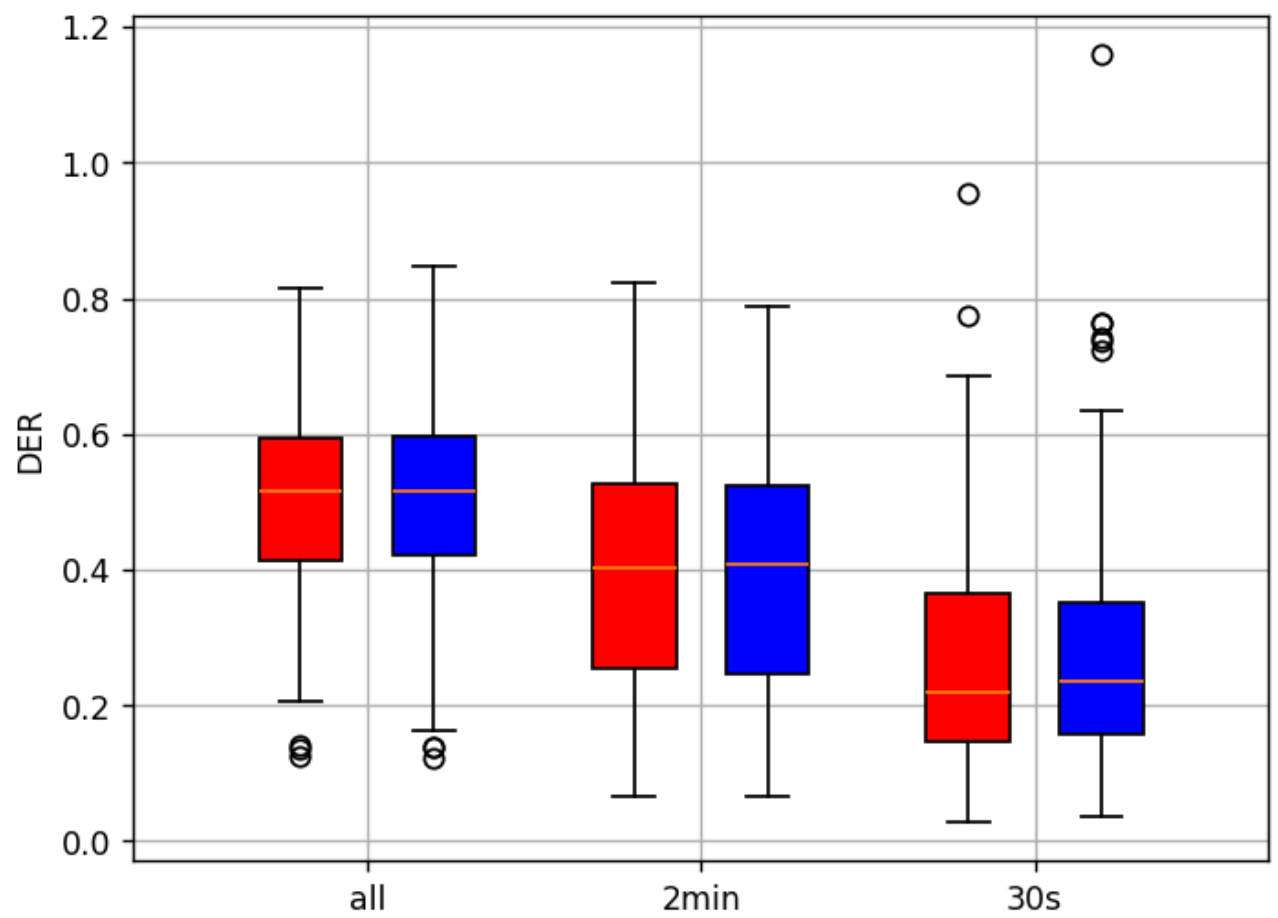

Figure 2: DER for the tuned model with VAD (left/red) and without VAD (right/blue). We measure the DER for the whole audio ("all"), the first 2 mins and the first 30 secs.

From Figure 1 a strong correlation between cpWER and DER is visible (r=0.76). The mean cpWER and DER for 2-speaker audios are 55% and 46%, while for 3-speaker audios it is 84.4% and 60.9%. In view of this metric difference it is important to note two properties regarding 3-speaker audios: their underrepresentation in the training set (Table 1) and the sporadicity with which we found the third speakers to appear throughout the test set audios (~20% of the whole speech). Due to sparsity the third speaker can be "pushed" out of the context prompt by the other (dominant) speakers, and their labelled text does not get propagated to the later audio chunks. Having not only text but also audio context should make the model more robust, so approaches like encoder prompting [3] sound like a potential solution.

Another reason for high DER is the propagation of error. As the model relies on prompts for getting the speaker label history, any mislabelling can cascade down the rest of the transcription. This is visible from Figure 2, where the model's median DER doubles after the first 30s of the audio. Word recognition mistakes can also impact the interpretability of the prompts, reflected in the moderate (r=0.43) correlation between WER and DER. Thus, we expect that by lowering overall WER we can lower DER (and consequently cpWER).

We also observed that the timestamps, estimated using Whisper's cross-attention scores, are inaccurate for overlapping speech. For the samples from the test subset we noticed that words from short overlapping segments get timestamped sequentially. For large overlapping segments the Faster Whisper word aligner assigns the same timestamp to every word from the segment. A possible solution is adjusting Whisper's attention-based alignment [12] or using an external aligner.

Lastly, we considered the tuned model's performance with and without VAD (Figure 2), thinking that the silence between speech could be of help in identifying speaker change. However, we found no significant difference between the DERs (two-sided Wilcoxon test p-value of 0.75). This can be another indication that the model relies more strongly on the text context for speaker labelling. Potentially, unmasking the prompt during training played a key role in enforcing this.

## 6. CONCLUSION

Fine-tuning Whisper with prompts yielded improved verbatim transcription and generated more uniform speaker labels. Though the model's ability to preserve speaker labels between audio chunks using prompting is promising, many challenges remain, namely correcting timestamps assigned to overlapping speech and providing audio context for more robust speaker distinction.

## ACKNOWLEDGEMENTS

This research was supported by the MediSpeech project funded by ITEA4 under contract number 22032.